\documentclass[global,twocolumn,referee]{svjour}
\usepackage{amsmath}

  \usepackage{graphics}

\journalname{myjournal}
\begin{document}

\title{Removal of surface plasmon polariton eigenmodes degeneracy}

\author{Xi-Feng Ren, Guo-Ping Guo\thanks{E-mail: \email:gpguo@ustc.edu.cn}, Pei Zhang, Yun-Feng Huang, Zhi-Wei Wang, Guang-Can Guo}

\institute{
  Key Laboratory of Quantum Information, University of Science and Technology of China, Hefei
230026, People's Republic of China }

\date{Received: date / Revised version: date}

\maketitle

\begin{abstract}
The effect of the tilt angle of metal film on the transmissivity of
subwavelength holes in optically thick metal film is investigated.
The transmission efficiency is found can be highly dependent on the
tilt angle. It is also found that when the input photons are
polarized not along the eigenmode directions of surface plasmon
polariton, a birefringent phenomenon is observed when the array with
periodic of subwavelength holes is tilted. Linear polarization
states can be changed to elliptical polarization states and a phase
can be added between two eigenmodes. The phase is changed with the
tilt angle. A model based on surface plasmon polariton eigenmodes
degeneracy is presented to explain these experimental results.
\end{abstract}

\section{Introduction}
Since it was first reported in 1998\cite{1}, the extraordinary
optical transmission (EOT) through periodic arrays of subwavelength
holes has attracted much attention owing to its fundamental
implications and its technological potential. Generally, it is
believed that metal surface plays a crucial role and the phenomenon
is mediated by surface plasmon polaritons (SPPs) and there is a
process of transform photon to surface plasmon polariton and back to
photon\cite{4,crucial,ebbesen5}. The polarization of the incident
light determines the mode of excited surface plasmon which is also
related to the periodic structure. This phenomenon can be used in
various applications, for example, sensors, opto electronic device,
etc\cite{williams,brolo,nahata,luo,shinada}.

In 2002, E. Altewischer \textit{et al.} \cite{alt} first showed that
polarization entanglement of photon pairs can be preserved when they
respectively travel through a hole array. Therefore, the macroscopic
surface plasmon polarizations, a collective excitation wave
involving typically $10^{10}$ free electrons propagating at the
surface of conducting matter, have a true quantum nature. After
that, polarization properties of EOT were investigated in many
works. For example, strong polarization dependence was observed in
EOT of elliptical nanoholes\cite{Elli04,Gordon04}; polarization
tomography of metallic nanohole arrays has been done
recently\cite{Altew05,Altew052}; and state of polarization was also
studied and modeled as a function of angular bandwidth in
EOT\cite{Altew053}. However, the increasing use of EOT requires
further understanding of the phenomenon.

For the manipulation of light at a subwavelength scale with periodic
arrays of holes, two ingredients exist: shape and
periodicity\cite{4,crucial,ebbesen5,klein,elliott,Ruan,sarra}. The
mode of excited surface plasmon polariton is related to the periodic
structure and the polarization of the incident light. Usually, the
surface plasmon polariton eigenmodes are degenerate due to the
symmetry of the hole array and the polarization of photon can be
preserved in the EOT process\cite{alt,Altew05}. In a recent paper,
splitting of degenerate surface plasmon polariton eigenmodes was
observed\cite{Lin}. Then, is there any influence on the polarization
of transmitted photons if the degeneracy of surface plasmon
polariton eigenmodes are removed? A feasible method to remove the
degeneracy is tilting the hole array. In our work, it was found that
photons with the same linear polarization can be transformed into
various elliptical polarization states if the hole array was tilted.
In detail, if the photons were polarized along the eigenmode
directions of surface plasmon polaritons, the polarization states
were not changed. Otherwise, a phase was generated between different
surface plasmon polaritons, as in a birefringent process. A linear
polarization state $cos\alpha_1 |H\rangle+sin\alpha_1|V\rangle$ was
changed to $cos\alpha_2 |H\rangle+sin\alpha_2 e^{i\beta}|V\rangle$.
The phase $\beta$ was varied with the tilt angle $\theta$.  This
phenomenon might come from the removal of degeneracy of surface
plasmon polariton eigenmodes.
\section{Experimental results}
\begin{figure}[b]
\resizebox{0.45\textwidth}{!}{%
  \includegraphics{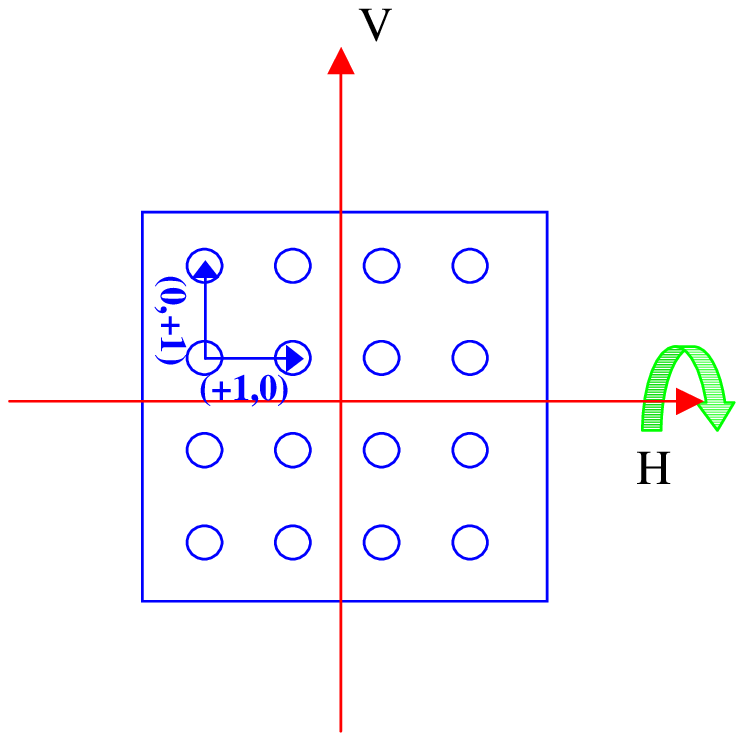}
} \caption{(color online). Sketch map of our hole array. The arrow
indicates the tilt direction. The thickness of the hole array is
about $140 nm$, and the holes have a diameter of $200 nm$ and period
of $600 nm$.}
\end{figure}

\begin{figure}[b]
\resizebox{0.45\textwidth}{!}{%
  \includegraphics{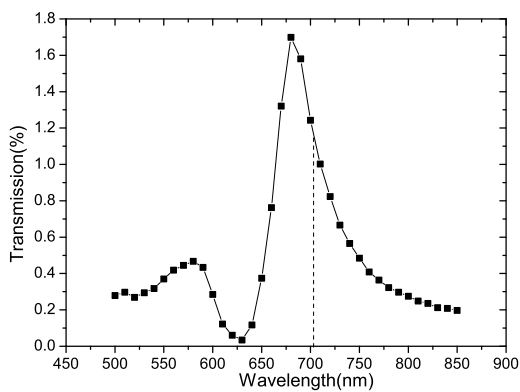}
}
\caption{(color online). Hole array transmittance as a function of
wavelength. The dashed vertical line indicates the wavelength of
$702 nm$ used in the experiment.}
\end{figure}

\begin{figure}[b]
\resizebox{0.45\textwidth}{!}{%
  \includegraphics{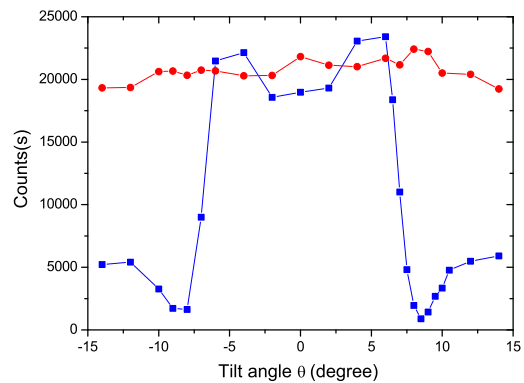}
} \caption{(color online). Relation between transmission
efficiencies for photons and tilt angle $\theta$. The red round dots
are the counts for horizontal polarized photons, and the blue square
dots are the counts for the vertical polarized photons. For photons
in horizontal polarization, the transmission efficiency was kept
constant, while for photons in vertical polarization, the
transmission efficiency was changed with the tilt angle $\theta$.}
\end{figure}

\begin{figure}[b]
\resizebox{0.45\textwidth}{!}{%
  \includegraphics{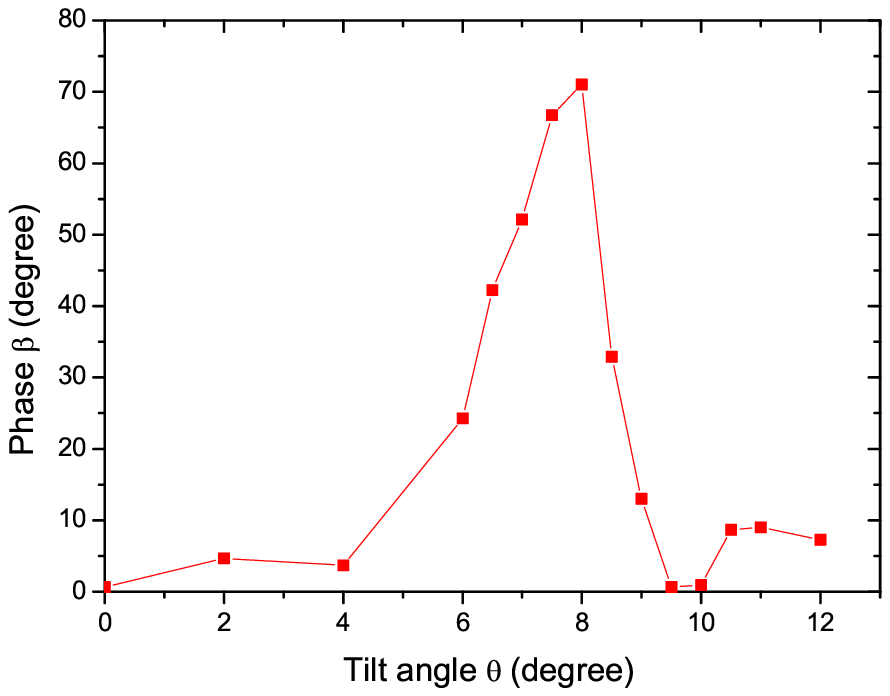}
} \caption{(color online). Relation between phase $\beta$ and tilt
angle $\theta$. For the linear polarization input state
$1/\sqrt{2}(|H\rangle+|V\rangle)$, the output state becomes an
elliptical polarization state
$cos(\alpha)|H\rangle+sin(\alpha)e^{i\beta}|V\rangle)$.}
\end{figure}

Fig. 1 shows the sketch map of our hole array. It is produced as
follows: after subsequently evaporating a $3$-$nm$ titanium bonding
layer and a $135$-$nm$ gold layer onto a $0.5$-$mm$-thick silica
glass substrate, a Electron Beam Lithography System (EBL, Raith 150
of Raith Co.) is used to produce cylindrical holes ($200nm$
diameter) arranged as a square lattice ($600nm$ period). The total
area of the hole array is $300\mu m\times 300\mu m$. The arrow
indicates the tilting direction. Transmission spectra of the hole
arrays for vertical polarized photons was shown in Fig. 2. The
dashed vertical line indicates the wavelength of $702 nm$ used in
our experiment. The transmission efficiency of the metal plate at
$702 nm$ is about $1.24\%$, which is much larger than the value of
$0.66\%$ obtained from classical theory\cite{Bethe}.

In experiment, white light from a stabilized tungsten-halogen source
passed though single mode fiber and $4 nm$ filter (center wavelength
702 nm) to generate 702nm wavelength photons. Polarization of input
light was controlled by a polarizer, a HWP (half wave plate, 702nm)
and a QWP (quarter wave plate, 702nm). The hole array was set
between two lenses of $35 mm$ focal length, so that the light was
normally incident on the hole array with a cross sectional diameter
about $20\mu m$ and covered hundreds of holes. Symmetrically, a QWP,
a HWP and a polarizer were combined to analyze the polarization of
transmitted photons. Silicon avalanche photodiode (APD) photon
counter was used to record counts(We use neutral density filter to
reduce the intensity of input light).

For our sample, photons with $702nm$ wavelength will excite the
surface plasmon polariton eigenmodes $(0,\pm 1)$ and $(\pm 1,
0)$(see Fig. 1). The hole array was tilted in the vertical direction
with tilt angle $\theta$ as shown in Fig. 1. First, transmission
efficiencies of photons in horizontal and vertical polarization were
measured. For photons in horizontal polarization, the transmission
efficiency was kept constant, while for photons in vertical
polarization, the transmission efficiency was changed with the tilt
angle $\theta$( see Fig. 3). Further, polarization state tomography
method was used to analyze the influence of tilt angle on the
polarization properties of the hole array. It was found that for
photons which were polarized along the eigenmode directions (here
horizontal and vertical), the polarization of transmitted photons
were not changed. When tilt angle $\theta$ was changed from
$-14\textordmasculine $ to $+14\textordmasculine$, the fidelities
between the polarization states of input and output photons for
horizontal polarized photons were all higher than $99.4\%$, and for
vertical polarized photons all higher than $96.8\%$. It was
interesting that even the transmission efficiency of vertical
polarized photons was varied with $\theta$, the polarization state
was not changed.

Then photons in polarization state $1/\sqrt{2}(|H\rangle+|V\rangle)$
illuminated the hole array and the polarization of transmitted
photons were analyzed using polarization state tomography. We found
that even the input state was a linear polarization state, the
output state was changed to an elliptical polarization state
$cos(\alpha)|H\rangle+sin(\alpha)e^{i\beta}|V\rangle$. A phase
$\beta$ was generated in the EOT process when the metal plate was
tilted. The relation between $\beta$ and tilt angle $\theta$ is
given in Fig. 4. The similar results were also observed when the
input state was $1/\sqrt{2}(|H\rangle+i*|V\rangle)$.

\section{Modeling}

\begin{figure}[b]
\resizebox{0.45\textwidth}{!}{%
  \includegraphics{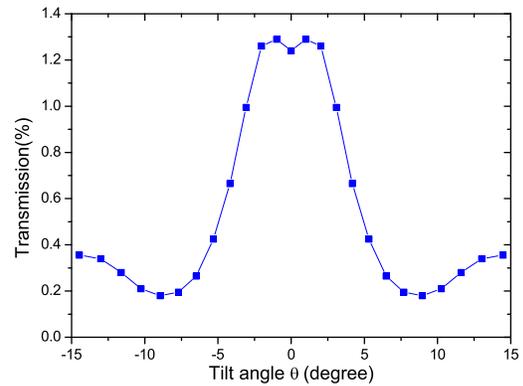}
} \caption{(color online). Theoretical calculated transmission
efficiencies for vertical polarized photons and tilt angle $\theta$.
It was just a simple addition of two moved transmission peaks.}
\end{figure}

A simple model was given below to explain the experimental results.
As we know, the interaction of the incident light with surface
plasmon polariton is made allowed by coupling through the grating
momentum and obeys conservation of momentum
\begin{equation}
\overrightarrow{k}_{sp}=\overrightarrow{k}_{0}\pm
i\overrightarrow{G}_{x}\pm j\overrightarrow{G}_{y},
\end{equation}
where $\overrightarrow{k}_{sp}$ is the surface plasmon wave vector,
$\overrightarrow{k}_{0}$ is the component of the incident wave
vector that lies in the plane of the sample,
$\overrightarrow{G}_{x}$ and $\overrightarrow{G}_{y}$ are the
reciprocal lattice vectors for a square lattice with
$\overrightarrow{G}_{x}=\overrightarrow{G}_{x}=2\pi/d$ , and i, j
are integers. For our sample, surface plasmon polariton eigenmodes
$(+1,0)$ and $(-1,0)$ are always degenerate when the hole array is
tilted vertically. If the input state is $|H\rangle$, the whole EOT
process can be described briefly as:
\begin{equation}
|H\rangle _{photon}\rightarrow (|+1, 0\rangle_{sp}+|-1,
0\rangle_{sp})\rightarrow |H\rangle _{photon}.
\end{equation}
Obviously, there will be no influence on the transmission of
horizontal polarized photons(including the transmission efficiency
and polarization) as we observed. While in vertical direction,
$\overrightarrow{k}_{sp(0,+1)}\neq \overrightarrow{k}_{sp(0,-1)}$
due to the nonzero $\overrightarrow{k}_{0}$. The degeneracy between
surface plasmon polariton eigenmodes $(0,+1)$ and $(0,-1)$ is
removed. In this case, momentum of the surface plasmon polariton,
photons and structure were also quasi-conserved. The reason was that
the hole diameter was $200nm$, which can not be neglected comparing
with the periodicity $600nm$. When the photons in vertical
polarization illuminate the hole array, two different surface
plasmon polariton eigenmodes are excited. The transmission peaks
corresponding these two eigenmodes will move away in opposite
direction(see Fig .3 of Ref[22]). Since the transmitted photons were
eradiated from the two surface plasmon polariton eigenmodes, the
whole transmission efficiency was the sum of the two cases. This
definitely influence the transmission efficiency of photons in 702nm
wavelength as shown in Fig. 3. In the theoretical calculation, the
whole transmission of Fig. 2 was divided into two equal
parts($(0,+1)$ mode and $(0,-1)$ mode), the two parts moved to short
wave and long wave directions respectively(see Fig .3 of Ref[22])
when the metal plate was tilted. A simple addition of two moved
transmission peaks was given in Fig 5, while in the calculation we
supposed that the shape of the transmission peaks was not changed
for the sake of simplification. It had a similar curve with the
experimental data.

Further, since $\overrightarrow{k}_{sp(0,+1)}\neq
\overrightarrow{k}_{sp(0,-1)}$, a phase is generated between these
two different photon to surface plasmon polariton and back to photon
process:

\begin{equation}
\begin{split}
|V\rangle _{photon}&\rightarrow (|0, +1\rangle_{sp}+|0,
-1\rangle_{sp})\\
&\rightarrow (|V\rangle _{photon_{|0,
+1\rangle}}+e^{i\varphi}|V\rangle _{photon_{|0, -1\rangle}}).
 \end{split}
\end{equation}
So if the input state is $1/\sqrt{2}(|H\rangle+|V\rangle)$, the
output state becomes
$cos(\alpha)|H\rangle+sin(\alpha)e^{i\beta}|V\rangle)$. Of course,
phase $\beta$ was correlated with the difference between
$\overrightarrow{k}_{sp(0,+1)}$ and
$\overrightarrow{k}_{sp(0,-1)}$(or $2\overrightarrow{k}_{0}$), which
was determined by the rotation angle $\theta$. $\beta$ will increase
for larger tilt angle $\theta$ in a small range. While when $\theta$
became bigger, the conservation of momentum was destroyed. The
analysis above will be not appropriate. Fig. 4 gives the relation
between phase $\beta$ and tilt angle $\theta$. The shape of the line
agreed with our prediction.
\section{Conclusion and discussion}
In conclusion, we found that when the hole array was tilted in the
EOT process, not only the transmission spectra was influenced, but
also a birefringent phenomenon appeared. Linear polarization state
$cos\alpha_1 |H\rangle+sin\alpha_1|V\rangle$ can be changed to
elliptical one: $cos\alpha_2 |H\rangle+sin\alpha_2
e^{i\beta}|V\rangle$. Since the phase $\beta$ can be controlled by
changing the tilt angle $\theta$, we can modulate the birefringent
phenomenon in EOT process. Using a simple model based on surface
plasmon polarition, we briefly explained the experimental results,
while further works are need to give a detailed explanation. Our
results may give us more hints on the polarization properties of
EOT.

This work was funded by the National Fundamental Research Program,
National Nature Science Foundation of China (10604052), Program for
New Century Excellent Talents in University, the Innovation Funds
from Chinese Academy of Sciences and the Program of the Education
Department of Anhui Province (Grant No.2006kj074A). Xi-Feng Ren also
thanks for the China Postdoctoral Science Foundation (20060400205)
and the K. C. Wong Education Foundation, Hong Kong.

\end{document}